\documentclass[11pt]{article}

\usepackage{geometry}  % Flexible and complete interface to document dimensions
\usepackage[T1]{fontenc}  % Standard package for selecting font encodings
\usepackage[utf8x]{inputenc}  % Accept different input encodings
\usepackage{lineno}  % Line numbers on paragraphs
\usepackage{fancyhdr}  % Extensive control of page headers and footers in LaTeX2ε
\usepackage{enumitem}  % Control layout of itemize, enumerate, description
\usepackage{hyperref}  % Extensive support for hypertext in LaTeX
\usepackage{titling}  % Control over the typesetting of the \maketitle command
\usepackage{natbib}  % Flexible bibliography support
\usepackage{mathtools}  % Mathematical tools to use with amsmath
\usepackage{titlesec}  % Select alternative section titles
\usepackage{multirow}
\usepackage{lastpage}  % Reference last page for Page N of M type footers
\usepackage[table]{xcolor}
\usepackage[english]{babel}  % Multilingual support for Plain TeX or LaTeX
\usepackage{amsmath}  % AMS mathematical facilities for LaTeX
\usepackage{amsfonts}  % TeX fonts from the American Mathematical Society
\usepackage{amssymb}  % Additional symbols from American Mathematical Society
\usepackage{wasysym}  % LaTeX support file to use the WASY2 fonts
\usepackage{bbm}  % "Blackboard-style" cm fonts
\usepackage{array}  % Extending the array and tabular environments
\usepackage{xr}  % References to other LaTeX documents
\usepackage{verbatim} % Reimplementation of and extensions to LaTeX verbatim
\usepackage{float} % Improved interface for floating objects
\usepackage{caption}
\usepackage{subcaption}
\usepackage[superscript,biblabel]{cite}
\usepackage{scalefnt}
\usepackage{adjustbox}

\hypersetup{%
  pdfborderstyle={/S/U/W 1}% border style will be underline of width 1pt
}

\setcitestyle{round,numbers}
\titleformat{name=\section}{\normalfont\Large\bfseries}{}{0pt}{}
\titleformat{name=\subsection}{\normalfont\large\bfseries}{}{0pt}{}
\titleformat{name=\subsubsection}{\normalfont\normalsize\bfseries}{}{0pt}{}

\newcommand{\email}[1]{\href{mailto:{#1}}{{#1}}}

\newcommand{\keywords}[1]{\textbf{Keywords}: {#1}}

\newcommand{\wordcount}[2]{\begin{tabular}{rl}%
\textbf{Manuscript word count}: 	& {#1}\\
\textbf{Abstract word count}: 		& {#2}\\
\end{tabular}}

\newcommand{\optincludegraphics}[2][]{}
\newcommand{\optinput}[1]{}
\newtagform{brackets}{[}{]}
\usetagform{brackets}

\newcommand{\thejournal}[1]{Magnetic Resonance in Medicine}

\title{Spectro-ViT: A Vision Transformer Model for GABA-edited MRS Reconstruction Using Spectrograms}
\begin{document}

% ======================================================================
%TC:ignore
\begin{titlepage}
{\noindent\LARGE\bf \thetitle}

\bigskip

\begin{flushleft}\large
	Gabriel Dias\textsuperscript{1},
        Rodrigo Pommot Berto\textsuperscript{2-4},
        Mateus Oliveira\textsuperscript{1},
	Lucas Ueda\textsuperscript{1,5},
        Sergio Dertkigil\textsuperscript{6},
        Paula D. P. Costa\textsuperscript{1,7},
        Amirmohammad Shamaei\textsuperscript{3,8},
	Roberto Souza\textsuperscript{3,8},
	Ashley Harris\textsuperscript{3,4,9},
        Leticia Rittner\textsuperscript{1}
\end{flushleft}

\bigskip

\noindent
%FIXME
\begin{enumerate}[label=\textbf{\arabic*}]
\item School of Electrical and Computer Engineering, University of Campinas, Campinas, Brazil
\item Department of Biomedical Engineering, University of Calgary, Calgary, Canada
\item Hotchkiss Brain Institute, University of Calgary, Calgary, Canada
\item Alberta Children’s Hospital Research Institute, Calgary, Canada
\item Research and Development Center in Telecommunications, CPQD, Campinas, Brazil
\item School of Medical Sciences, University of Campinas, Campinas, Brazil
\item Artificial Intelligence Lab., Recod.ai, University of Campinas, Campinas, Brazil
\item Department of Electrical and Software Engineering, University of Calgary, Calgary, Canada
\item Department of Radiology, University of Calgary, Calgary, Canada

% Not sure where this came from - Talal had Libin support but not Stepehen?
%\item Libin Cardiovascular Institute, University of Calgary, Calgary, Alberta
\end{enumerate}

\bigskip

% : Use the dagger symbol to denote a single equal contribution authorship.
% : Multiple equal-contribution authorship may be included in the acknowledgments.
%FIXME
%\textbf{{†}}: These authors contributed equally to this work.

% : Use the asterisk to denote corresponding authorship.
% : Provide email address in note below.
%FIXME
\textbf{*} Corresponding author:

\indent\indent
\begin{tabular}{>{\bfseries}rl}
Name		&  Roberto Souza, PhD													\\
Department	& Electrical and Software Engineering													\\
Institution	& University of Calgary														\\
Address 	 & ICT 352C \\
            & 2500 University Dr NW													\\
			& Calgary, Alberta														\\
            & Canada	T2N 1N4  													\\
E-mail		& \email{roberto.souza2@ucalgary.ca}											\\
\end{tabular}

\vfill
\wordcount{4691}{197}
% ======================================================================
% : set word count results (+++ must be included, --- must be excluded)
% 	+++ introduction, theory, methods, results, discussion, conclusion,
%		appendix, 
% 	--- title page, abstract, figure captions, tables, table captions,
%		references, revision markings
% : first argument is the manuscript word count
% : second argument is the abstract word count
% : to use `texcount` results, use '%TC:ignore'/'%TC:endignore' directives.
% : \wcManuscript and \wcAbstract should perform the correct word count.

% : display detailed word count
%FIXME

\end{titlepage}
%TC:endignore
% ======================================================================

% ======================================================================
% ======================================================================
\pagebreak
% ======================================================================
% ======================================================================

% ======================================================================
%TC:break Abstract
\begin{abstract}

\textbf{Purpose}: To investigate the use of a Vision Transformer (ViT) to reconstruct/denoise GABA-edited magnetic resonance spectroscopy (MRS) from a  quarter of the typically acquired number of transients using spectrograms.

\textbf{Theory and Methods}: A quarter of the typically acquired number of transients collected in GABA-edited MRS scans are pre-processed and converted to a spectrogram image representation using the Short-Time Fourier Transform (STFT). The image representation of the data allows the adaptation of a pre-trained ViT for reconstructing GABA-edited MRS spectra (Spectro-ViT).  The \textit{Spectro-ViT} is fine-tuned and then tested using \textit{in vivo} GABA-edited MRS data. The \textit{Spectro-ViT} performance is compared against other models in the literature using spectral quality metrics and estimated metabolite concentration values.

\textbf{Results}: The \textit{Spectro-ViT} model significantly outperformed all other models in four out of five quantitative metrics (mean squared error, shape score, GABA+/water fit error, and full width at half maximum). The metabolite concentrations estimated (GABA+/water, GABA+/Cr, and Glx/water) were consistent with the metabolite concentrations estimated using typical GABA-edited MRS scans reconstructed with the full amount of typically collected transients.

\textbf{Conclusion}: The proposed \textit{Spectro-ViT} model achieved state-of-the-art results in reconstructing GABA-edited MRS, and the results indicate these scans could be up to four times faster.

\end{abstract}

% ======================================================================
% : set search-engine keywords (3 to 6)
\bigskip

%TC:break _main_
% ======================================================================
% ======================================================================
\pagebreak
% ======================================================================
% ======================================================================

% ======================================================================
\section{Introduction}
% ======================================================================
Gamma-aminobutyric acid (GABA) is the primary inhibitory neurotransmitter in the mammalian brain and plays a crucial role in healthy brain function\cite{mccormick1989gaba}. Understanding GABAergic mechanisms is vital for both normal and pathological brain states, representing a key area in neuroscience\cite{mikkelsen2017big}. The quantification of GABA level changes in patients has primarily been investigated using magnetic resonance spectroscopy (MRS) \cite{peek2023comprehensive,hasler2007reduced}, a non-invasive technique that allows for the identification and measurement of brain metabolites \cite{oz2014clinical,tomiyasu2022vivo,lin2012guidelines,cecil2013proton}.

The assessment of GABA abnormalities levels through MRS has been observed in diverse neuropsychiatric disorders\cite{mikkelsen2017big}. These disorders encompass conditions like schizophrenia \cite{kegeles2012elevated,ongur2010elevated,rowland2016medial} and depression \cite{bhagwagar2008low}, neurodevelopmental disorders, such as autism spectrum disorder \cite{puts2017reduced,drenthen2016altered} and attention deficit hyperactivity disorder \cite{edden2012reduced,bollmann2015developmental}, as well as neurological diseases including Parkinson's disease\cite{emir2012elevated}, amyotrophic lateral sclerosis\cite{foerster2013imbalance}, and diabetic neuropathy \cite{petrou2012altered}.

The most widely used MRS sequence to measure GABA is the J-difference editing \cite{star1998vivolactate,mescher1998simultaneous,choi2021spectral} implementation known as MEscher–GArwood Point RESolved Spectroscopy (MEGA-PRESS)\cite{mescher1998simultaneous,mikkelsen2017big}. MEGA-PRESS has gained popularity due to several reasons, including its extensive availability across different scanner platforms \cite{mikkelsen2017big}, its relatively simple implementation \cite{mullins2014current}, its reproducibility \cite{shungu2016brain,mikkelsen2016comparison,brix2017within}, and ongoing advancements in acquisition techniques and data processing tools\cite{chan2016hermes,edden2014gannet}.

MEGA-PRESS involves obtaining two sub-signals: one uses specific radiofrequency pulses (RF) to invert the GABA signal (Edit-ON), while the other lacks inversion, allowing regular evolution of the spin system throughout the echo time (Edit-OFF). Subtracting the sub-signal ON from the sub-signal OFF isolates GABA by subtracting overlapping metabolites and eliminating stronger signals from creatine-containing compounds.

Although the separability of the GABA signal is significantly improved using MEGA-PRESS, the method is technically challenging due to a notable limitation: the time required to obtain high-quality data for accurate detection and quantification. This challenge is intricately linked to the process of signal averaging \cite{de2019vivo}, which is a fundamental aspect of MEGA-PRESS. Signal averaging involves acquiring repeated measurements of the ongoing experiment to enhance the signal-to-noise ratio (SNR). The obtained signals from these repetitions are then averaged at the end of the scan. 
%The temporal issue stems from the fact that SNR increases only with the square root of the number of signals accumulated for signal averaging \cite{ernst1966sensitivity,traficante1991time}. 
%The duration of the scan time in MRS is directly linked to signal averaging. 
Achieving a significant SNR improvement can lead to impractical prolongation of scan time, as SNR increases only with the square root of the number of transients \cite{ernst1966sensitivity,traficante1991time,de2019vivo}. 
%For example, obtaining a tenfold improvement in SNR would require a hundredfold increase in scan time. This trade-off between desired signal quality and scan duration constraints becomes evident when dealing with low SNR signals.

In the literature, these repeated signals for signal averaging are commonly referred to by different terms, such as transients, averages, excitations, or acquisitions  \cite{peek2023comprehensive}. Extensive discussions can be found in the literature regarding the optimal number of transients needed to achieve good signal quality \cite{mullins2014current, mikkelsen2017big,mikkelsen2019big,harris2014impact,mikkelsen2018designing,brix2017within,sanaei2018quantification,bhattacharyya2007spectral, peek2023comprehensive}. The recommended range varies from 126 (i.e. 63 Edit-ON + 63 Edit-OFF)  \cite{sanaei2018quantification} to the highest recommended value of 320 (i.e. 160 Edit-ON + 160 Edit-OFF)  \cite{mikkelsen2017big, mikkelsen2019big}. 
Current acquisition times can last over 10 minutes considering the recommendation of 320 transients\cite{peek2023comprehensive}.

In order to enhance the clinical applicability of MRS, it is crucial to maximize the  data quality while minimizing lengthy scan times. The development and adoption of techniques that could reduce the scan time would not only enhance the feasibility and practicality of MRS in clinical applications but would also decrease costs and improve patient scheduling efficiency.

%Reconstruction is a commonly used term in the literature, describing the process of recovering information lost through acceleration techniques. It corresponds to a relevant field that studies ways to speed up the technique while maintaining data quality.  In the context of MRI images, Deep Learning (DL) has offered promising solutions for reconstruction. DL-based reconstruction is now a fundamental field of MRI research that is rapidly progressing, with numerous methods already proposed~\cite{sun2016deep,wang2016accelerating,schlemper2017deep,schlemper2017deep,chen2018variable,knoll2019assessment,mardani2018deep,lundervold2019overview,beauferris2022multi}. Within this context, recent studies have also shown promising evidence supporting the utilization of DL models in MRS\cite{nassirpour2018multinet,lee2020reconstruction,iqbal2021deep,motyka2021k, chan2022improved, van2023review,berto2023investigation}. 

Medical imaging applications have witnessed a growing utilization of Deep Learning (DL) to enhance the techniques owing to its adaptability and capability to manage complex data. DL techniques have found application in the enhancement of data quality and the acceleration of conventional MRS data acquisition by mitigating noise and reducing scan duration\cite{lee2019intact,chen2023magnetic,wang2023denoising,berto2023investigation}. In the realm of enhancing the MEGA-PRESS, researchers in the field have proposed DL solutions to tackle the limitations commonly associated with traditional pre-processing techniques for frequency and phase correction (FPC) \cite{tapper2021frequency,ma2022mr,shamaei2023model}. 
%These limitations encompass issues such as poor performance at lower SNRs and extended computational time for handling large datasets.

The initial FPC study based on DL by Tapper et al.\cite{tapper2021frequency} introduced a multilayer perceptron (MLP) model with frequency domain input that outperformed conventional non-DL-based methods in medium to large datasets. Ma et al.\cite{ma2022mr} enhanced this MLP by incorporating a convolutional neural network (CNN), marking the first utilization of CNNs for FPC. This CNN exceeded the previous MLP in terms of performance and noise resilience. Alternatively, Shamaei et al \cite{shamaei2023model} employed an unsupervised deep autoencoder with time domain input for FPC. When compared to five FPC techniques, including the previous DL-based solutions \cite{tapper2021frequency,ma2022mr}, this method exhibited improved precision and introduced the similarity index for measuring alignment quality. It also excelled in linewidth analysis, even when dealing with nuisance peaks, while demonstrating faster processing times than traditional limited frequency methods.

Despite these efforts, these proposed methods do not resolve the trade-off between signal quality and scan time. While FPC helps improve signal quality by minimizing line broadening and subtraction artifacts\cite{harris2014impact,choi2021spectral}, the improvement in SNR is inherently related to the amount of collected data (transients)\cite{de2019vivo}. As a result, the acquisition time when using these FPC models remains unchanged.

In the context of reducing MEGA-PRESS scan times while maintaining data quality, Berto et al.'s investigation \cite{berto2023investigation} stands out as first work in the literature that addressed the trade-off inherent in signal averaging. Berto et al.\cite{berto2023investigation} introduced a DL model capable of reconstructing MEGA-PRESS scans using only 80 transients (40 Edit-ON + 40 Edit-OFF), achieving higher SNR and lower fit error compared to the conventional reconstruction pipeline, which typically uses the full 320 transients. The study also examined different training schemes, including using only \textit{in-vivo} data or pre-training the models on simulated data and then fine-tuning them with \textit{in-vivo} data. The results indicated that the Rdc-UNET2D model, pre-trained on simulated data and fine-tuned with \textit{in-vivo} data, outperformed other implemented DL pipelines by the same study. 

A subsequent study in the literature adopted the same  investigation of reconstructing MEGA-PRESS scans using only 80 transients as part of the Edited-MRS reconstruction challenge \cite{berto2023advancing}. This challenge featured four distinct DL models, each exploring different representations of typically one-dimensional spectroscopy data in a two-dimensional format. The findings demonstrated that the leading DL-based edited-MRS reconstruction pipelines achieved metrics comparable to traditional methods using 320 transients, despite only utilizing 80 transients. Additionally, the study introduced a novel metric known as the `shape score,' which exhibited a positive correlation with the challenge outcomes, indicating its effectiveness in assessing spectral quality. The model presented in this work secured the first position in the challenge.

In continuation of recent research efforts in the literature \cite{berto2023investigation,berto2023advancing}, this study delves deeper into the potential of DL for reducing scan time in MEGA-PRESS for GABA measurement, introducing an enhanced state-of-the-art DL model that secured the first-place solution in the edited-MRS reconstruction challenge \cite{berto2023advancing}. The contributions of this paper are as follows:

\begin{itemize}

%The adoption of the proposed DL reconstruction model could reduce the acquisition time to 2.5 minutes.
 
%\item This work aim to address the gap in the literature concerning signal averaging reconstruction for MEGA-PRESS. The proposed DL model is expected to establish a benchmark for future research, setting a new standard in efficient and accurate reconstruction of GABA-edited MRS scans. 

\item 

\textbf{State-of-the-art model:} A state-of-the-art model that addresses the trade-off in signal averaging for MEGA-PRESS in GABA measurement. The proposed DL model has achieved better reconstruction metrics than the one developed by Berto et al~\cite{berto2023investigation}. It also secured the first-place solution in the Edited-MRS reconstruction challenge \cite{berto2023advancing} for reconstructing GABA MEGA-PRESS scans using only 80 transients.

\item \textbf{Exploration of the spectrogram domain:} the successful use of features extracted from the spectrogram domain to train a DL reconstruction model, showcasing a promising alternative for MRS related tasks.  While previous models in the MEGA-PRESS literature typically chose either a frequency or time domain representation as input, the proposed approach explores the spectrogram's ability to balance these two aspects.

\item \textbf{Vision Transform (ViT):}
The use a ViT model \cite{dosovitskiy2020image} to solve a MRS problem. A recent literature review \cite{van2023review}  provides a comprehensive overview of machine learning applications in MRS, revealing a lack of Transformer models in this domain. The review emphasizes the importance of exploring different and novel model types, including Transformers, which have demonstrated potential in other medical imaging fields \cite{li2023transforming}. Therefore, this work pioneers the use of the Transformer model, marking its first application in MRS research. By leveraging the capabilities of the ViT, this study aims to unlock new possibilities and insights in MRS analysis, potentially enhancing its performance and opening methods for further advancements.

\item \textbf{Quantification Investigation:} This study presents the first investigation of spectra quantification analysis in GABA MEGA-PRESS DL-based reconstruction and goes beyond reconstruction metrics. Statistical analysis and data visualization demonstrate that the proposed model by this work yields metabolite concentration estimation comparable to those of the conventional pipeline, which uses 320 transients.

\item \textbf{Transients Subset Analysis:} In this study, 320 transients were analyzed  by dividing them into four groups, each representing a quarter of the data acquisition process. This division allows to assess the robustness of the innovative ViT model in addressing common challenges in MEGA-PRESS spectral analysis, such as frequency and phase misalignment caused by subject movement \cite{mullins2014current}. The model's stability was evaluated by analyzing transients 1-80, 81-160, 161-240, and 241-320, which simulated different levels of spectral quality. The consistent results across these groups demonstrated the method's robustness.

\item \textbf{Share of code and model:} 
By openly share its findings and code, this work intends to facilitate easy reuse and to encourage further similar research in the future. \url{https://github.com/MICLab-Unicamp/Edited_MRS_Challenge}

\end{itemize}

% ======================================================================
% ======================================================================

% ======================================================================
\section{Methods}

\subsection{Dataset}
\label{sec:data}

This study utilized \textit{in-vivo} data sourced from the Big GABA repository \cite{mikkelsen2017big,mikkelsen2019big,povavzan2020comparison}.This dataset comprises 144 standard GABA+ edited MEGA-PRESS scans, each with 320 transients, obtained from GE, Philips, and Siemens scanners across twelve sites: G4, G5, G7, G8, P4, P6, P8, P10, S1, S3, S5, and S8. The number of scanners per vendor was the same, equally divided, totaling 48 scans per vendor. Each site was instructed to comply with the standardized protocol, maintaining a fixed voxel size of 30x30x30 mm$^3$ in the medial parietal lobe while also avoiding ventricles and/or the outer surfaces of the brain when necessary to ensure good data quality. The acquisition parameters are detailed in Mikkelsen et al.\cite{mikkelsen2017big,mikkelsen2019big}. Furthermore, the subjects in the dataset had no reported neurological or psychiatric illnesses, with a mean age of 26.4 ± 4.1 years. These data correspond to the same \textit{in-vivo} data used by Berto et al. \cite{berto2023investigation}.

Following the same experimental setup of Berto et al. \cite{berto2023investigation}, this research divided the scans into three subsets: 84 for training, 24 for validation, and 36 for testing. Importantly, each subset maintained an equal number of scans per site, and the age distribution of subjects in the modelling subsets was stratified. The extraction of the 160 Edit-On and 160 Edit-Off Free Induction Decay (FID) signals 
%, which are used to compute the spectrogram, the input of the model, 
was performed using Gannet \cite{edden2014gannet}, a widely recognized MATLAB-based open-source software for processing, fitting and quantifying GABA-edited spectra within the MRS community \cite{harris2017edited}.

% \begin{table}[h]
% \caption{Distribution of MEGA-PRESS scans across training, validation, and test subsets, with equal representation from GE, Philips, and Siemens scanners. This table shows the count of scans per vendor and the mean age of subjects, with standard deviation, for each subset used in the study. Data were sourced from the Big GABA repository\cite{mikkelsen2017big,mikkelsen2019big,povavzan2020comparison}.}

% \label{table:modeling_split}
% \centering
% \begin{tabular}{|c|c|c|c|c|}
% \hline
% \rowcolor[HTML]{EFEFEF} 
% \textbf{Set} &
%   \textbf{\begin{tabular}[c]{@{}c@{}}Philips \\ Count\end{tabular}} &
%   \textbf{\begin{tabular}[c]{@{}c@{}}Siemens \\ Count\end{tabular}} &
%   \textbf{\begin{tabular}[c]{@{}c@{}}GE \\ Count\end{tabular}} &
%   \textbf{Age Mean \(\pm \sigma\)} \\ \hline
% Train          & 28 & 28 & 28 & 25.797  ± 3.782                               \\ \hline
% Validation     & 8  & 8  & 8  & \cellcolor[HTML]{FFFFFF}26.250 ± 4.445        \\ \hline
% Test           & 12 & 12 & 12 & \cellcolor[HTML]{FFFFFF}28.027 ± 4.266        \\ \hline
% \textbf{Total} & 48 & 48 & 48 & \multicolumn{1}{l|}{\cellcolor[HTML]{EFEFEF}} \\ \hline
% \end{tabular}
% \end{table}

A method for accommodating the model's input limited to 80 transients, while working with \textit{in-vivo} data consisting of 320 transients, was implemented, enabling the effective utilization of the data without discarding the excess 240 transients. The method uses a sliding window with a size of 40, which moves along the transient axis, creating groups of 40 transients per sub-signal (ON and OFF). This technique not only addresses the data limitation but also enhances the model's learning robustness across different transients, which may exhibit varying levels of spectral quality.

% \begin{figure}[H]
%   \centering
%   \includegraphics[width=12cm]{figs/sliding window method.png}
%   \caption{The sampling method involves a window that slides along the transient axis with a unitary step, forming groups of 40 transients per sub-signal. For 320 transients (160 per sub-signal), the algorithm iterates 120 times, resulting in the creation of 120 groups (samples).}
%   \label{fig:sliding_window}
% \end{figure}

As a result, this method generated an additional 120 samples, each defined as a set of 40 transients per sub-signal (Edit-ON and Edit-OFF). These samples are subsequently used to generate the spectrogram input for the model. After applying the sliding window method to both the training and validation sets, a total of 10,080 samples (120$\times$84) were generated for training, and a total of 2880 samples (120$\times$24) were generated for validation purposes.

% \begin{figure}[H]
%   \centering
%     \includegraphics[width=\textwidth]{figs/spectra_distribution.png}
%   \caption{Visualization depicting MEGA-PRESS target spectra distribution in training, validation, and test subsets, with respective sample sizes in parentheses. It displays the mean edited spectrum signal in blue with ±1 standard deviation shading in orange. Individual spectra were normalized using maximum absolute scaling prior to mean signal computation.}\label{fig:spectra_distribution}
% \end{figure}

\begin{figure}[H]
  \centering
  \includegraphics[width=0.9\textwidth]{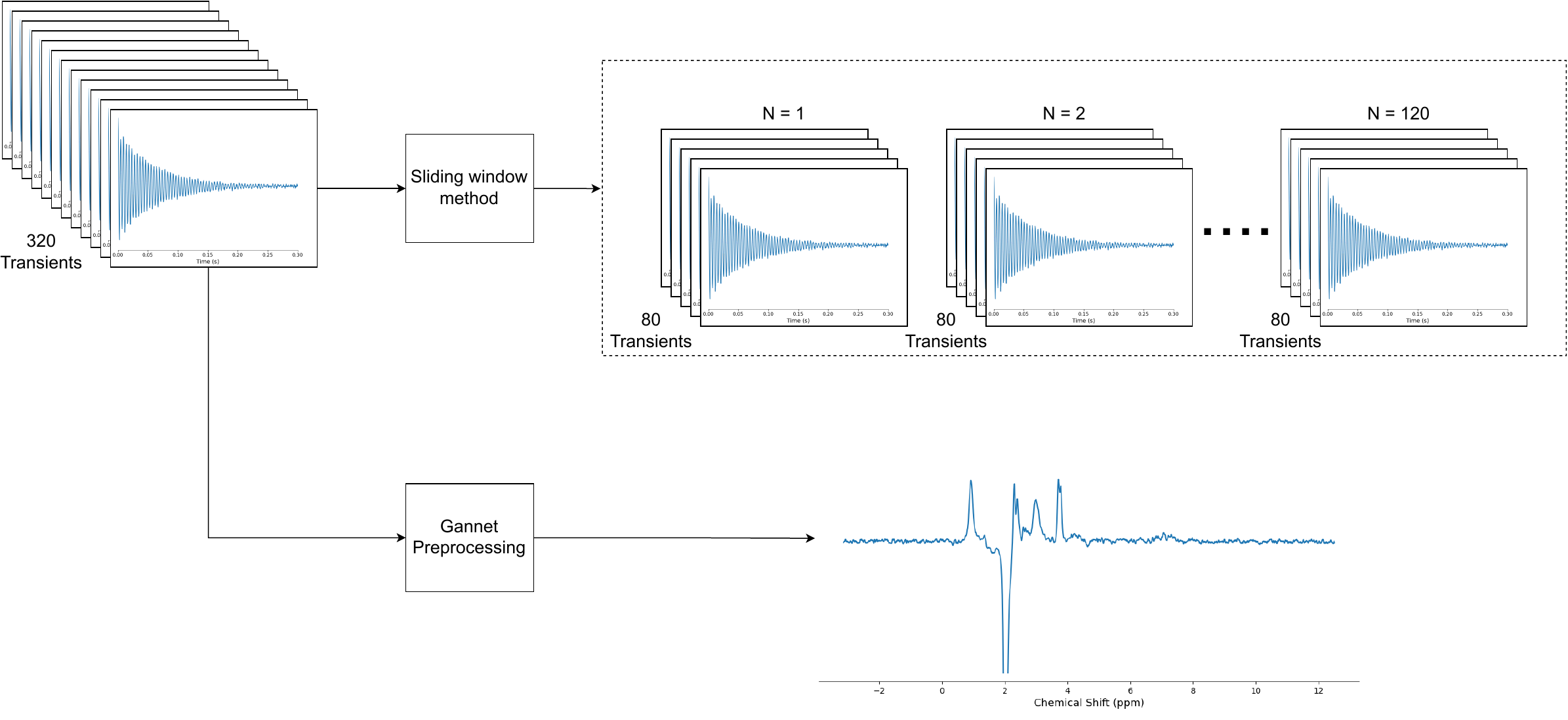}
  \caption{The input data for training the model, consisting of 80 transients, was generated using the sliding window method from the complete set of 320 \textit{in-vivo} transients. The target data were obtained from the preprocessing pipeline, which utilized the full set of transients and the Gannet software. In this process, the 120 samples generated by the sliding window method shared the same target spectrum.}
  \label{fig:in_vivo_generate}
\end{figure}

The target data for each sample corresponded to the edited spectrum produced by the Gannet preprocessing pipeline (v.3.3.2) with default settings, which utilized 320 transients \cite{edden2014gannet} (Fig.~\ref{fig:in_vivo_generate}). The pipeline included eddy current correction applied to the unsuppressed water signal \cite{klose1990vivo}, line-broadening (apodization) using a 3 Hz weighting constant, robust spectral registration for both frequency and phase alignment \cite{near2015frequency, mikkelsen2020correcting}, and zero-filling. Signal averaging employed the weighted averaging method, reducing the contribution of lower-quality transients to the final edited spectra.

After processing the Gannet data, down-sampling was performed in the edited spectrum using nearest neighbour interpolation to ensure that all spectra had a fixed number of data points, specifically 2048 data points. This step eliminated excessive data points from frequencies that did not significantly contribute to the overall information, thereby highlighting the relevant features within the spectra, improving computational storage efficiency, and simplifying the architecture's modelling. Also, the target spectra were normalized by dividing it by its maximum absolute value. One target-edited spectrum was generated for each scan, with the 120 samples generated by the sliding window method sharing the same target spectrum (Fig. \ref{fig:in_vivo_generate}).

\subsection{Spectrogram Generation}
\label{sec:spectrogram}

Spectrogram is a common visual representation of the Short-time Fourier transform (STFT) \cite{hlawatsch1992linear}, a technique used to analyze signals in the time-frequency domain. It involves dividing a time-domain signal into segments and computing the Fourier Transform (FT) for each time segment~\cite{KEHTARNAVAZ2008175}. This allows the identification of frequency information that is specific to particular time intervals, which is especially useful when the frequency components of a signal change over time. In contrast to the traditional FT, which provides frequency information averaged over the entire signal duration, the STFT provides time-localized frequency information. The STFT's accuracy and resolution are determined by parameters such as window size, hop size, sampling rate, window type, and FT length, parameters chosen based on the desired analysis.

Spectrograms can be seen as one-channel images, where the x-axis represents time frames, y-axis the frequencies and each pixel represents the magnitude of each frequency component at each time frame. Thus, it is intuitive to use vision models on spectrogram to perform different kinds of tasks. In audio domain this approach is already widely used for processing tasks~\cite{ke2005computer,mariotti2018exploring,le2019using}. Some clinical areas that work with different types of biological signals have also already shown the effectiveness of using this representation in classification tasks~\cite{diker2019a, aslan2020automatic}.

\begin{figure}[!htbp]
  \centering
  \includegraphics[width=\textwidth]{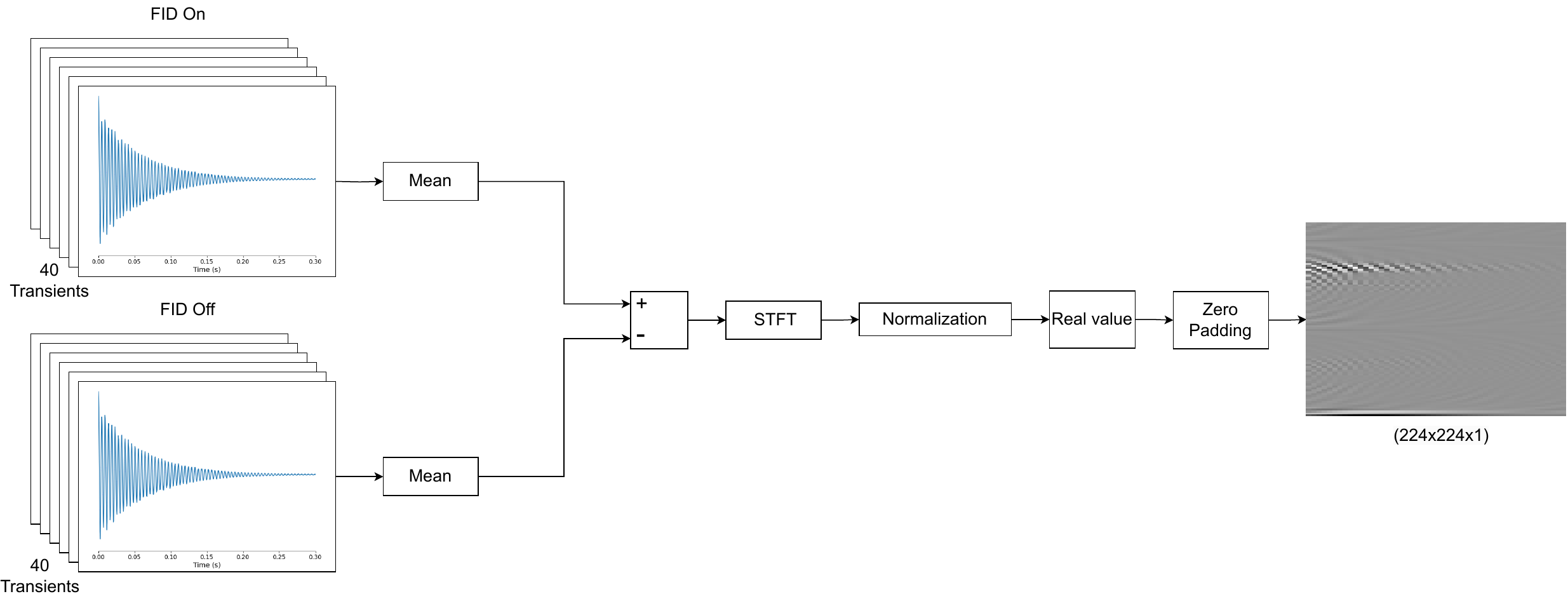}
  \caption{Spectrogram generation for each sample involves averaging 40 Edit-ON and 40 Edit-OFF FID transients, followed by subtracting the mean Edit-ON FID from the mean Edit-OFF FID. The Short-Time Fourier Transform (STFT) is then computed on the resulting signal, with spectrogram normalization and zero-padding enhancing compatibility for Vision Transformer Model (ViT) integration at a size of $224 \times 224$. The output is a 1-channel image (gray-level) spectrogram.}
  \label{fig:spectrogram_generation}
\end{figure}

In order to generate the spectrogram for each sample, the following steps were performed (Fig.~\ref{fig:spectrogram_generation}):

\begin{enumerate}
    \item The average was calculated across the 40 Edit-ON FID transients, as well as for the 40 Edit-OFF FID transients;
    \item The mean Edit-ON FID was subtracted from the mean Edit-OFF FID;
    \item The STFT was computed for the resulted FID signal. For signals with 2048 data points, the window size, hop size, and FFT length were set to 256, 10, and 446, respectively. For signals with 4096 points, only the hop size was changed to 20;
    \item The spectrogram was normalized by dividing it by its maximum magnitude value.
    \item The real values from the spectrogram were computed.
    \item Zero-padding was performed to increase the spectrogram size to $224 \times 224$, preparing it for further use into the Vision Transformer Model, which was pre-trained on $224\times224$ images.
\end{enumerate}

\begin{figure}[!htbp]
  \centering
    \includegraphics[width=\textwidth]{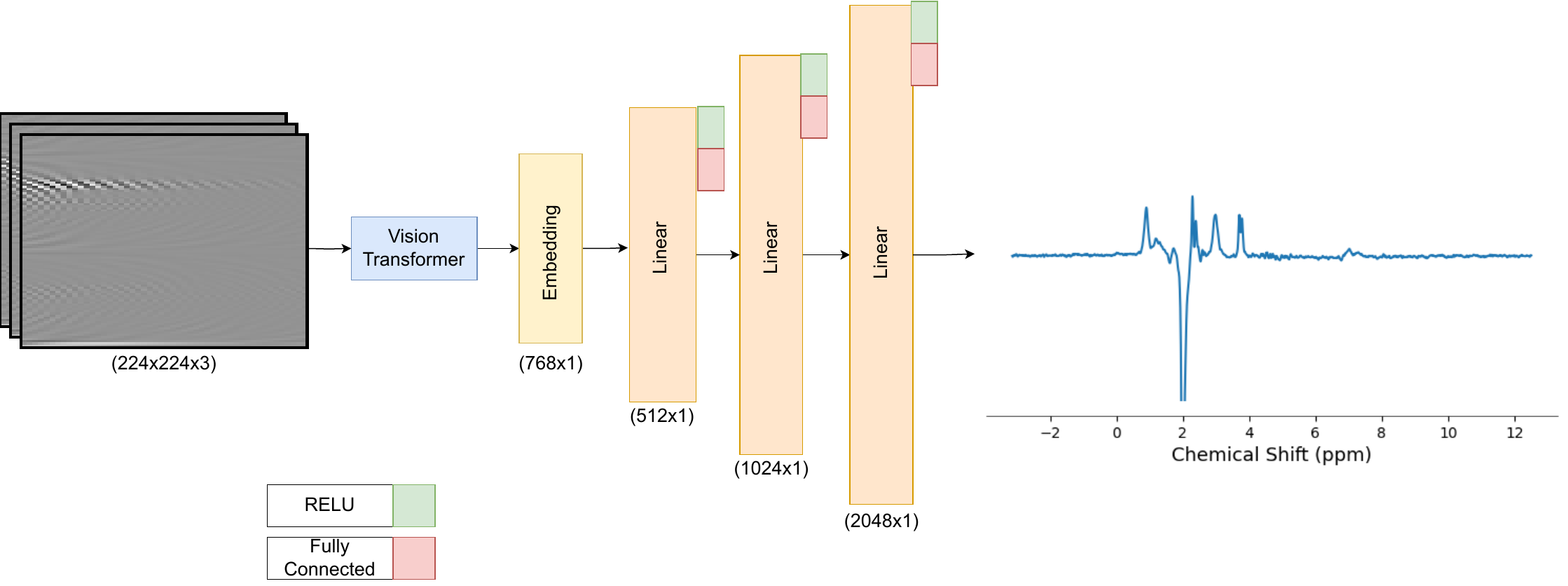}
  \caption{The ViT receives the real-valued spectrogram as input with 3 replicated channels and outputs a 768-dimensional embedding, that passes through fully connected layers with ReLU activation ($512\times1$, $1024\times1$, and $2048\times1$). This final layer generates a GABA-edited spectrum with 2,048 data points.}\label{fig:model_architecture}
\end{figure}

\subsection{The Vision Transformer (ViT) Model}
\label{sec:vit}

The Vision Transformer (ViT) model, introduced by Dosovitskiy et al.~\cite{dosovitskiy2020image}, represents a significant advancement in computer vision by applying transformer architectures, traditionally used in Natural Language Processing (NLP), to this field. Distinct from conventional Convolutional Neural Networks (CNNs), the ViT is a part of the broader category of Transformers, which utilize attention, residual, and memory techniques for more effective data modeling. Transformers have demonstrated impressive performance in various computer vision domains, including image classification \cite{dosovitskiy2020image}, video classification \cite{arnab2021vivit}, and  medical image segmentation \cite{chen2021transunet,li2021medical}.

The ViT marks a significant advancement in computer vision by its ability to prioritize the most important regions of input data, unlike traditional models that focus on local neighborhood relations in early layers. This is achieved through several key steps: dividing the input image into smaller, fixed-size patches for more effective processing; transforming each patch into an embedded vector representation; and employing a self-attention mechanism on these embeddings \cite{dosovitskiy2020image}. This self-attention, a hallmark of the Transformer encoder, allows the ViT to discern and interpret the relationships and interdependencies among different segments of the image, leading to more insightful image analysis.

A key component of this study is the combination of a pre-trained ViT model and a spectrogram. Transfer learning from vision tasks to audio tasks involving spectrograms has been previously studied with good results for CNN based models\cite{gwardys2014deep,guzhov2021esresnet}. Gong et al. \cite{gong2021ast} introduced the \textit{Audio Spectrogram Transformer} (AST), the first convolution-free, purely attention-based model for audio classification, taking the spectrogram as input. The AST model was adapted from the ViT architecture pretrained on ImageNet \cite{deng2009imagenet}. The AST achieved new state-of-the-art results on three different metrics across audio classification benchmarks, demonstrating its high performance in extracting features from spectrograms.

In this research, a pre-trained ViT model on ImageNet \cite{deng2009imagenet} was also used. The ViT model, which processes images at a 224x224 pixel resolution, breaks down images into 32x32 pixel patches. The ViT model is fed with a real-valued three-channel spectrogram (Fig.\ref{fig:spectrogram_generation}), producing a 768-dimensional embedding. This embedding subsequently progresses through fully connected layers, resulting in a 2048-dimensional output. This final layer contributes to the spectral reconstruction, resulting in a GABA-edited spectrum containing 2048 data points (Fig.\ref{fig:model_architecture}).

\subsection{Training}
\label{sec:train}

The model was trained in a supervised manner. Each sample (80 transients) gave origin to a spectrogram (Fig.\ref{fig:spectrogram_generation}), which is replicated across three channels and inputted into the model architecture (Fig.\ref{fig:model_architecture}). The model's output corresponds to the reconstructed MEGA-PRESS signal, which is compared with its corresponding target signal (Fig.\ref{fig:in_vivo_generate}) to compute the training loss, which is supposed to be minimized to enable the model's learning.

The model was trained with a batch size of 100, a learning rate of 0.0001, using the Adam optimizer, and for 50 epochs. During training, the dataset was shuffled once at the start of each epoch. Additionally, data corruption was applied to each sample individually as it was loaded into the model. This corruption involved a time-domain corruption by introducing random noise to the frequency, phase, and amplitude of the transients. This corruption process utilized a script from the GitHub repository of the Edited-MRS reconstruction challenge\cite{berto2023advancing}. The values of the noisy function parameters used in the script were uniformly sampled as follows: amplitude noise ranged from 0 (no noise) to 5 for the base level (applied to all transients and scans) and from 0 to 3 for scan variance (indicating the level of variation between different scans); frequency noise values ranged from 0 to 4 for the base level and from 0 to 3 for scan variance, while phase noise values ranged from 0 to 3 for both the base level and scan variance. This method enhances the algorithm's learning in realistic noisy scenarios and ensures that samples loaded in the same batch, generated from the same scan using the sliding window method, exhibit weaker correlations, thereby making them non-redundant.

The training loss employed was the same used in the study by Berto et al. \cite{berto2023investigation} (Eq.\ref{eq: loss}). This loss is a combined mean absolute error (MAE) approach, specifically targeting key spectral regions to prioritize the most significant attributes in the GABA-difference spectrum, namely the GABA peak at 3 ppm and the combined Glutamate and Glutamine (Glx) peak at 3.75 ppm.

\begin{equation}
\label{eq: loss}
\text{Loss} = \frac{(\text{MAE}_{\text{global}} + 3 \times \text{MAE}_{3.55-3.95ppm} + 6 \times \text{MAE}_{2.8-3.2ppm})}{10}
\end{equation}

\subsection {Evaluation Metrics}
\label{sec:metrics}

As the focus of this work is the GABA metabolite, all the evaluation metrics employed have been specifically designed to assess the quality of the edited reconstructed MEGA-PRESS spectrum for GABA measurement. These metrics have been previously utilized in research and encompass both a DL metric, Mean Squared Error (MSE), and three traditional MRS metrics: SNR \cite{lin2021minimum, choi2021spectral, near2021preprocessing}, Linewidth \cite{lin2021minimum, choi2021spectral, near2021preprocessing}, and Fit Error \cite{lin2021minimum, edden2014gannet}. Additionally, a recently proposed metric  from the literature  was included, known as Shape Score. The computation of these metrics is as follows:

- MSE: Measured between 2.5 ppm and 4 ppm, a min-max normalization was applied to both the reference and output pipeline spectra to ensure fair evaluation \cite{berto2023advancing}.

- SNR: Calculated by dividing the GABA peak height by twice the standard deviation of the 10 ppm to 12 ppm region.

- Linewidth: Determined as the Full Width at Half Maximum (FWHM), which denotes the width of the spectral peak at half of its maximum intensity. The FWHM was computed specifically over the GABA peak.

- Fit Error: Utilizing Gannet Software, the fit error is calculated during the fitting of signal functions to metabolite peaks. It is defined as the ratio of the signal model amplitude to the standard deviation of the model fit residuals. Importantly, the fit error of interest combines the errors from both the metabolite and the reference signal model fits, which are summed in quadrature \cite{edden2014gannet}.

- Shape Score: The shape score is a quantitative metric proposed by Berto et al. \cite{berto2023advancing}. It compares the normalized target with the normalized model's reconstruction, focusing on GABA and Glx peaks. It is calculated computing Pearson's correlation coefficient and weighting it as 60\% for GABA and 40\% for Glx. The signals are normalized using a min-max approach for each peak. The ppm range for normalizing the GABA peak is 2.8 - 3.2 ppm, while for the Glx peak, it is 3.55 - 3.9 ppm.

\section{Experiments}

Following the model training phase, a series of evaluation experiments were conducted on the test set as outlined below:

\subsection{1. Reconstruction Quality}

The first experiment was to evaluate the reconstruction in the same manner as reported in the literature in previous studies \cite{berto2023investigation,berto2023advancing}. The results of the proposed model \textit{Spectro-ViT} were compared with three other pipelines: the reconstructions by Gannet when using all the transients (\textit{Gannet Full}), when using only a quarter of the transients (\textit{Gannet quarter}), and with the DL model \textit{Rdc-UNET2D}~\cite{berto2023investigation}.

The edited spectra resulting from each of the pipelines were assessed using the metrics MSE, SNR, Linewidth, Shape Score, and GABA+/Water Fit Error, as well as visual inspection. Since each of the test samples contains 320 transients, for the pipelines that use only 80 transients (e.g., \textit{Spectro-ViT}, \textit{Gannet Quarter}, and \textit{Rdc-UNET2D}), the first 80 transients were used as input for these pipelines.

\subsection{2. Metabolite Quantification}

The second experiment involved spectra quantification analysis using the same pipelines from the previous experiment (\textit{Spectro-ViT}, \textit{Rdc-UNET2D}, \textit{Gannet Quarter}, \textit{Gannet Full}). This study presents the first investigation of spectra quantification analysis in GABA MEGA-PRESS DL-based reconstruction. The GABA+/Cr, GABA+/Water and Glx/Water measurements was performed using the Gannet Software \cite{edden2014gannet}.
    
For pipelines that employed a quarter of the conventional number of transients (\textit{Spectro-ViT}, \textit{Rdc-UNET2D}, \textit{Gannet Quarter}), the first 80 transients from the test samples were used. Since there are no ground-truth concentration values for the \textit{in-vivo} data, the \textit{Gannet Full} pipeline served as a reference for calculating the error for these measurements, specifically, the distribution of the mean absolute error and the mean absolute percentage error.

Furthermore, a Wilcoxon signed-rank test \cite{conover1999practical}  was conducted with the \textit{Gannet Full} pipeline as the reference. The null hypothesis tested whether the sample from each pipeline came from the same distribution as the reference, with a significance level of $\alpha=0.05$. This significance test was chosen because it compares paired data and makes no assumptions about the underlying distribution.

\subsection{3. Robustness Evaluation}

Finally, an innovative experiment for the existing literature involved the analysis of the reconstruction of a subset of transients. The robustness of the proposed \textit{Spectro-ViT} model was evaluated by analyzing the edited spectra resulting from the pipeline for different subsets of transients within the same sample. These groups were created by dividing the 320 transients into four distinct groups with no overlap: transients 1-80 (Group 1), 81-160 (Group 2), 161-240 (Group 3), and 241-320 (Group 4). Group 1 corresponds to the group used in experiments 1 and 2. The same reconstruction metrics from experiment 1 (GABA+/Water Fit Error, Linewidth, Shape Score, and SNR) were computed for the spectra generated for each group, and their distributions were compared.

Regarding quantification, in each group, the distributions and fit errors of GABA+/Water, Glx/Water, and GABA+/Cr were obtained with the Gannet Software \cite{edden2014gannet} and analyzed. Also, a comparison of the concentration values with the reference Gannet Full was made for each group using the Wilcoxon signed-rank test.

\section{Results}

\begin{figure}[!htbp]
  \centering
    \includegraphics[width=0.9\textwidth]{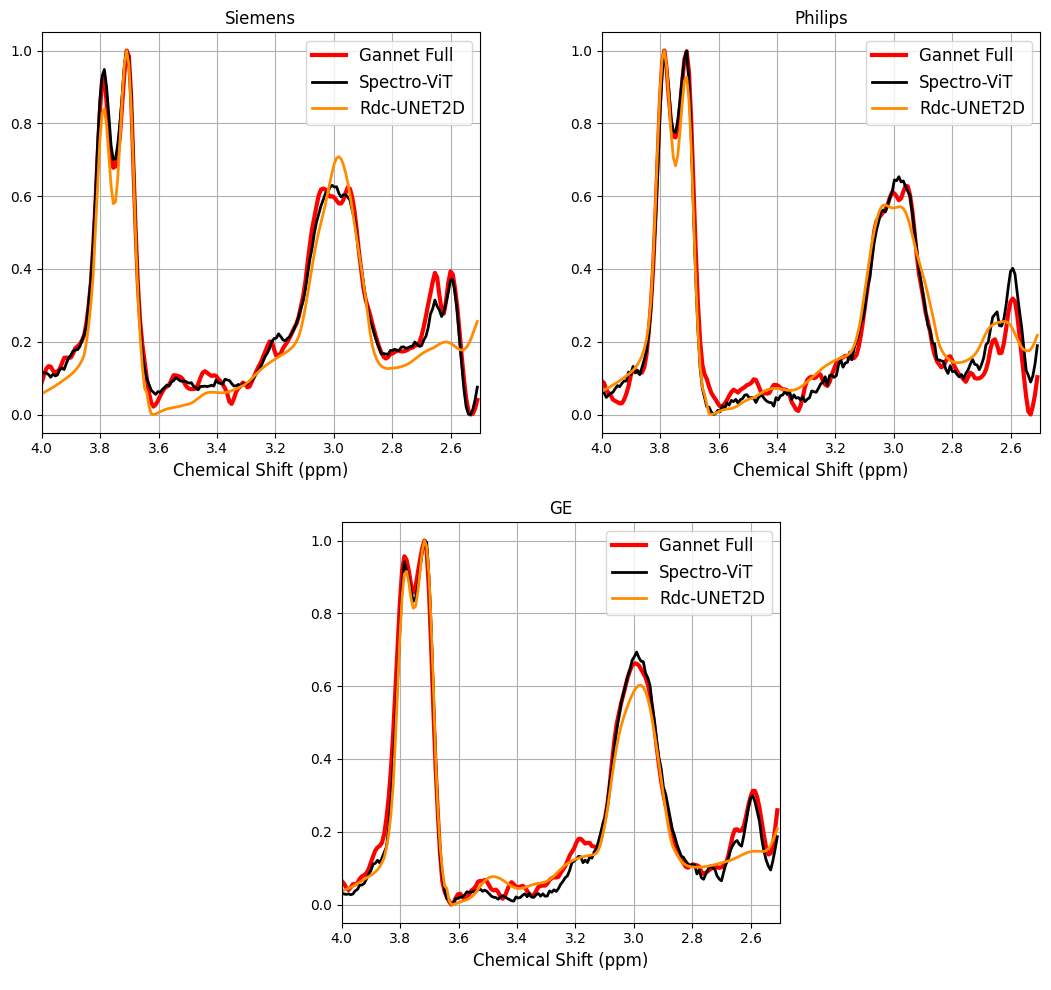}
  \caption{The test set's edited spectra, shown in the 2.5 to 4 ppm range, which is the same range used for the computation of the MSE metric evaluation (Tab. \ref{table:metrics}), normalized within this range. Spectra from Siemens, Philips, and GE scanners are presented. The DL pipelines \textit{Spectro-ViT} (black) and \textit{Rdc-UNET2D} (yellow) reconstructed spectra from the first 80 of 320 transients. The reference \textit{Gannet-Full} pipeline (red), using 320 transients, is also displayed. Both DL pipelines closely match the reference, with \textit{Spectro-ViT} more accurately representing GABA's peak height.}
  \label{fig:mixed_reconstruction}
\end{figure}

\textbf{1) Evaluation of Reconstruction:} \textit{Spectro-ViT} has demonstrated its superiority in most of the reconstruction metrics (Tab.\ref{table:metrics}). Specifically, the proposed model achieved the best mean shape score, mean MSE, fit error, and linewidth. The superiority of these two metrics can be visually observed in the reconstructed spectra (Fig.\ref{fig:mixed_reconstruction}). When analyzing the \textit{Spectro-ViT} results, it tends to approximate the reconstructed signal to the target better, particularly in capturing the height and shape of the GABA and Glx peaks. \textit{Spectro-ViT} stands out with the GABA+/Water Fit Error metric having the lowest mean, outperforming even the \textit{Gannet Full}, with a value approximately 2.6 times better and also values more consistent with the lowest standard deviation.

\begin{table}[htbp]
\footnotesize
\caption{Mean ± standard deviation of the metrics MSE, SNR, Linewidth (FWHM), Shape Score, and GABA+/Water Fit Error assessed over the edited spectra resulting from the pipelines \textit{Gannet Full}, \textit{Gannet Quarter}, \textit{Rdc-UNET2D}, and \textit{Spectro-ViT} (the proposed model) for the test set considering 320 transients for the \textit{Gannet Full} pipeline and the first 80 transients for the others. Best results for each metric are highlighted in bold.}
\label{table:metrics}
%\hspace*{-1.5cm}
\begin{tabular}{|c|c|c|c|c|c|}
\hline
\rowcolor[HTML]{EFEFEF} 
\textbf{Pipeline} &
  \textbf{\begin{tabular}[|c|]{@{}c@{}}MSE\\ (2.5 - 4 ppm)\end{tabular}} &
  \textbf{SNR} &
  \textbf{\begin{tabular}[|c|]{@{}c@{}}FWHM\\ (ppm)\end{tabular}} &
  \textbf{Shape Score} &
  \textbf{\begin{tabular}[|c|]{@{}c@{}}GABA+/Water \\ Fit Error \\ (\%)\end{tabular}} \\ \hline
Spectro-ViT    & \textbf{0.0087 ± 0.0154} & 45.1 ± 11.5          & \textbf{0.163 ± 0.015} & \textbf{0.983 ± 0.019} & \textbf{3.056 ± 1.139} \\ \hline
Rdc-UNET2D     & 0.0105 ± 0.0113          & \textbf{45.3 ± 21.5} & 0.166 ± 0.016          & 0.980 ± 0.016          & 4.394 ± 2.382          \\ \hline
Gannet Quarter & 0.0090 ± 0.0268          & 10.0 ± 3.1           & 0.182 ± 0.030          & 0.936 ± 0.261          & 7.971 ± 5.184          \\ \hline
Gannet Full    & N/A                      & 19.6 ± 5.5           & 0.168 ± 0.013          & N/A                    & 5.068 ± 2.477          \\ \hline
\end{tabular}
\end{table}

Regarding SNR, the key metric in signal averaging, it is worth noting that deep learning models can significantly improve this metric compared to the conventional \textit{Gannet Full} pipeline. While the \textit{Gannet Full} pipeline had an average SNR of 19.6, DL pipelines were able to more than double the value of the metric. Although the \textit{Spectro-ViT} model did not outperform the \textit{Rdc-UNET2D} model in terms of SNR, the mean difference was less than 0.4\%. Furthermore, the \textit{Spectro-ViT} model exhibited a considerably lower standard deviation in the same SNR metric  (Tab.\ref{table:metrics}).

\textbf{2) Spectra Quantification:}

In discussing the quantification results, the Gannet Quarter pipeline yielded the best results for absolute error and MAPE across all the studied relations. When comparing only DL-based pipelines, the \textit{Spectro-ViT} model achieved the best error results for GABA concentrations, while \textit{Rdc-UNET2D} performed best for Glx (Tab.\ref{table:quantification}). 

\begin{table}[!h]
\caption{Quantification results computed over the test set considering 320 transients for the \textit{Gannet Full} pipeline and the first 80 transients for the others. Distribution of absolute errors (mean ± standard deviation), mean absolute percentage error (MAPE), and Wilcoxon signed-rank test p-values across the pipelines \textit{Gannet Full}, \textit{Gannet Quarter}, \textit{Rdc-UNET2D}, and \textit{Spectro-ViT} for the quantification values of GABA+/Water, GABA+/Cr, and Glx/Water. The values obtained by the \textit{Gannet Full} pipeline were used as a reference for the calculation of these statistics. The `*' symbol assigned to the side of the p-value indicates a value below the significance level of 0.05.}

\label{table:quantification}
\centering
\begin{tabular}{|c|c|c|c|c|}
\hline
\rowcolor[HTML]{EFEFEF} 
\textbf{Metabolite} &
  \textbf{Pipeline} &
  \textbf{\begin{tabular}[c]{@{}c@{}}Absolute Error \\ Mean ± STD\end{tabular}} &
  \textbf{\begin{tabular}[c]{@{}c@{}}MAPE \\ (\%)\end{tabular}} &
  \textbf{\begin{tabular}[c]{@{}c@{}}Wilcoxon\\  p-value\end{tabular}} \\ \hline
\rowcolor[HTML]{FFFFFF} 
\cellcolor[HTML]{FFFFFF}                              & Spectro-ViT            & 0.2322 ± 0.1943          & 11.5129          & 0.9073   \\ \cline{2-5} 
\rowcolor[HTML]{FFFFFF} 
\cellcolor[HTML]{FFFFFF}                              & Rdc-UNET2D     & 0.2605 ± 0.2728          & 13.3162          & *0.0013 \\ \cline{2-5} 
\rowcolor[HTML]{FFFFFF} 
\multirow{-3}{*}{\cellcolor[HTML]{FFFFFF}GABA+/Water} & Gannet Quarter & \textbf{0.1933 ± 0.1818} & \textbf{10.5654} & *0.0442 \\ \hline
\rowcolor[HTML]{FFFFFF} 
\cellcolor[HTML]{FFFFFF}                              & Spectro-ViT            & 0.0140 ± 0.0149          & 11.5455          & 0.8584   \\ \cline{2-5} 
\rowcolor[HTML]{FFFFFF} 
\cellcolor[HTML]{FFFFFF}                              & Rdc-UNET2D     & 0.0160 ± 0.0210          & 13.5265          & *0.0017 \\ \cline{2-5} 
\rowcolor[HTML]{FFFFFF} 
\multirow{-3}{*}{\cellcolor[HTML]{FFFFFF}GABA+/Cr}    & Gannet Quarter & \textbf{0.0126 ± 0.0144} & \textbf{10.7907} & 0.0621   \\ \hline
\rowcolor[HTML]{FFFFFF} 
\cellcolor[HTML]{FFFFFF}                              & Spectro-ViT            & 0.8281 ± 0.6729          & 12.1099          & 0.6357   \\ \cline{2-5} 
\rowcolor[HTML]{FFFFFF} 
\cellcolor[HTML]{FFFFFF}                              & Rdc-UNET2D     & 0.7432 ± 0.6402          & 11.1914          & 0.7037   \\ \cline{2-5} 
\rowcolor[HTML]{FFFFFF} 
\multirow{-3}{*}{\cellcolor[HTML]{FFFFFF}Glx/Water}   & Gannet Quarter & \textbf{0.5385 ± 0.6918} & \textbf{7.9911}  & 0.9567   \\ \hline
\end{tabular}
\end{table}

\begin{figure}[htbp]
  \centering
    \includegraphics[width=0.95\textwidth]{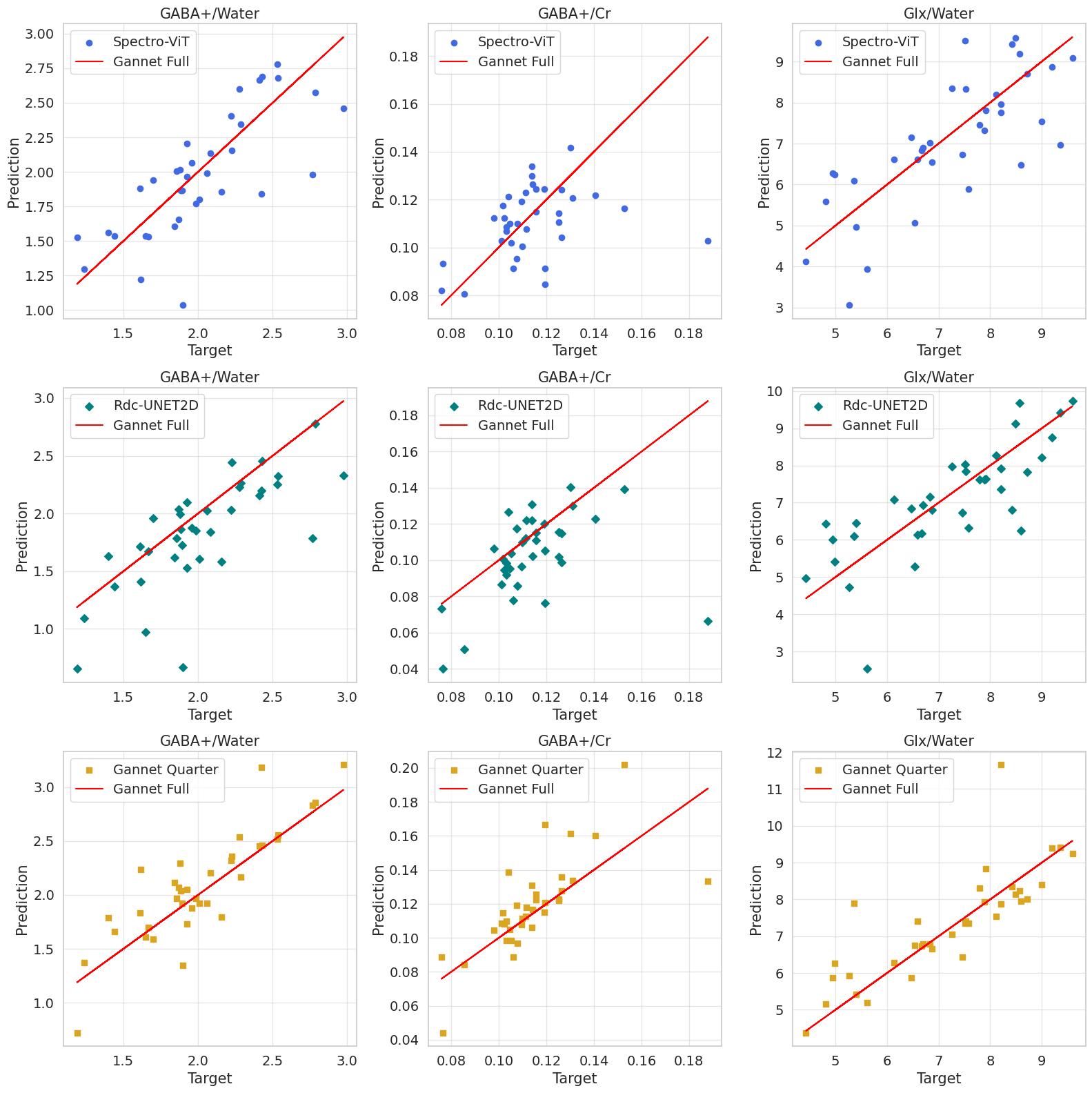}
  \caption{Scatter plots displaying concentration distributions computed over the test set are presented for three pipelines: \textit{Spectro-ViT} (proposed model) in the first row (in blue), \textit{Rdc-UNET2D} in the second row (in green), and \textit{Gannet Quarter} in the third row (in yellow). The values for these three pipelines (\textit{Spectro-ViT}, \textit{Rdc-UNET2D}, and \textit{Gannet Quarter}) were obtained by considering the first 80 transients of the test set. In all the plots, a red target line represents the values computed by the \textit{Gannet Full} pipeline using 320 transients of the test set. \textit{Spectro-ViT}'s GABA concentrations exhibit an even distribution around the \textit{Gannet Full} values, while \textit{Rdc-UNET2D} tends to underestimate, and \textit{Gannet Quarter} tends to overestimate, resulting in shifts in the distributions. All three pipelines demonstrate a balanced distribution of Glx/Water without apparent bias.}\label{fig:scatter_3v3_plot}
\end{figure}

However, when comparing the distributions to the \textit{Full Gannet}, \textit{Spectro-ViT} is the only pipeline that does not show a statistically significant difference to the target pipeline in every relation. \textit{Rdc-UNET2D} has considerably low p-values for GABA+/Water (0.0013) and GABA+Cr (0.0017), whereas \textit{Gannet Quarter} has a slightly lower p-value than 0.05 for GABA+/Water (0.0442) and a slightly higher value for GABA+/Cr (0.0621). All the pipelines generated high p-values for Glx/Water (Tab.\ref{table:quantification}). 
Also, the \textit{Spectro-ViT} had the lowest coefficient of variation for both GABA+/Water (22.3\%) and GABA+/Cr (13\%), while \textit{Gannet Quarter} scored 24.2\% and 22.6\%, and \textit{Rdc-UNET2D} scored 26\% and 22.2\%, respectively.

The scatter plots of the concentrations provide complementary information to the statistics (Fig.\ref{fig:scatter_3v3_plot}). GABA concentrations obtained through \textit{Spectro-ViT}'s reconstructed spectra are more evenly distributed around the \textit{Gannet Full} values, without displaying a bias toward one side of the reference red line. When examining \textit{Rdc-UNET2D} data, most values are close to the reference line, but the model tends to concentrate values and underestimate the target values. On the other hand, \textit{Gannet Quarter} tends to overestimate them, resulting in shifts in the distributions and statistical differences. For Glx/Water, all pipelines exhibit a balanced distribution without bias towards one side, with \textit{Gannet Quarter} having more values closer to the target line, consistent with the error metrics.

\textbf{3) Subset Robustness:} 

\begin{figure}[!htbp]
  \centering
    \includegraphics[width=\textwidth]{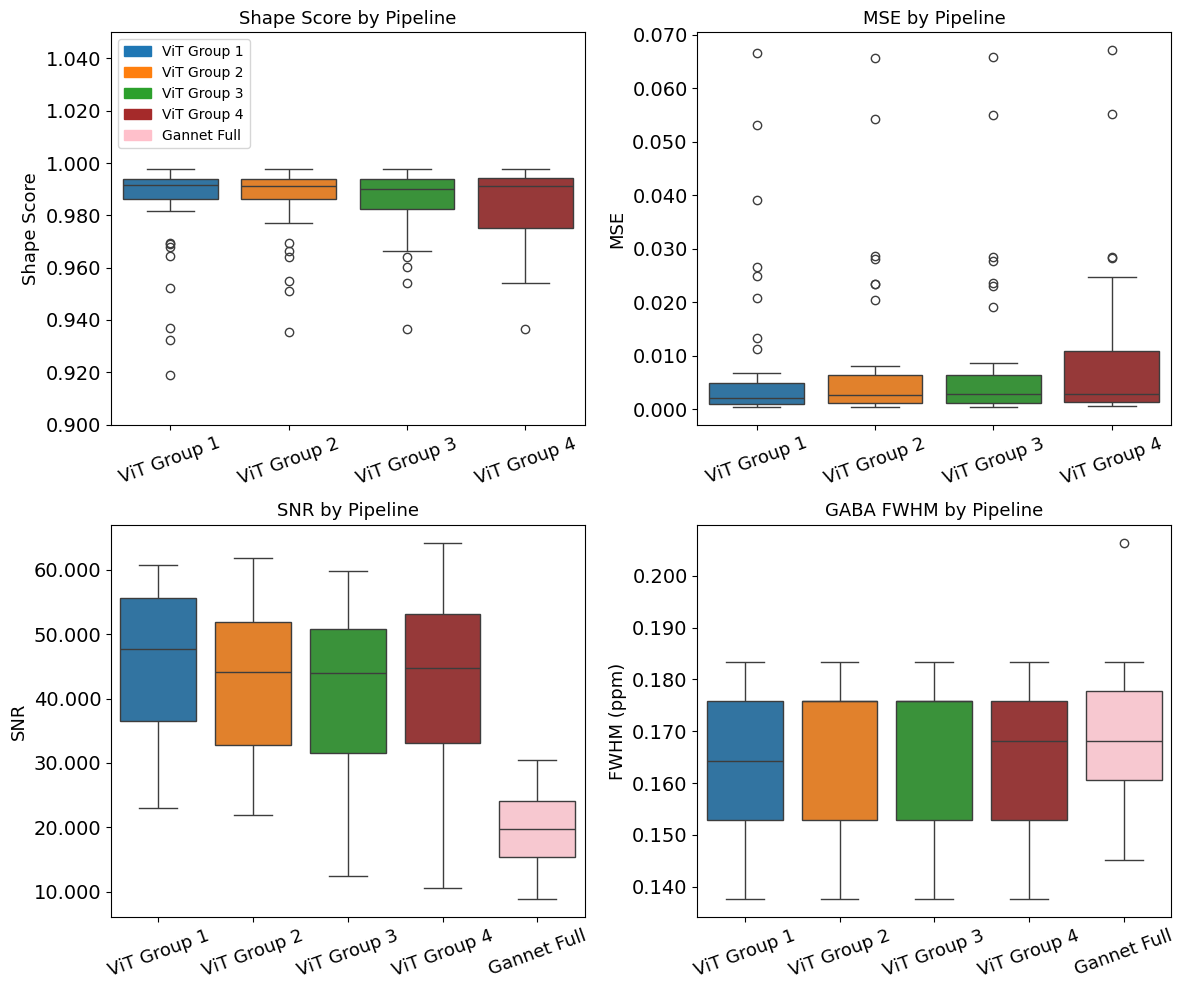}
  \caption{Box plots of metrics computed over the spectra resulting from the \textit{Spectro-VIT} pipeline (proposed model) for different subsets of transients in the test set. The \textit{Gannet-Full} pipeline, computed with 320 transients in the test set, is also displayed for reference in the SNR and FWHM metrics. Transients 1-80 (ViT Group 1) are shown in blue, 81-160 (Group 2) in orange, 161-240 (Group 3) in green, 241-320 (Group 4) in red, and \textit{Gannet-Full} pipeline in pink. In general, the \textit{Spectro-ViT} results demonstrate consistency with good values across the different subsets when considering the quality of reconstruction. The last transients have shown a higher susceptibility to frequency shifts, with Group 4 displaying the worst values for MSE and more dispersed values for Shape Score and SNR.}
  \label{fig:metrics_boxplot_sliced}
  
\end{figure}
\begin{figure}[!htbp]
  \centering
    \includegraphics[width=\textwidth]{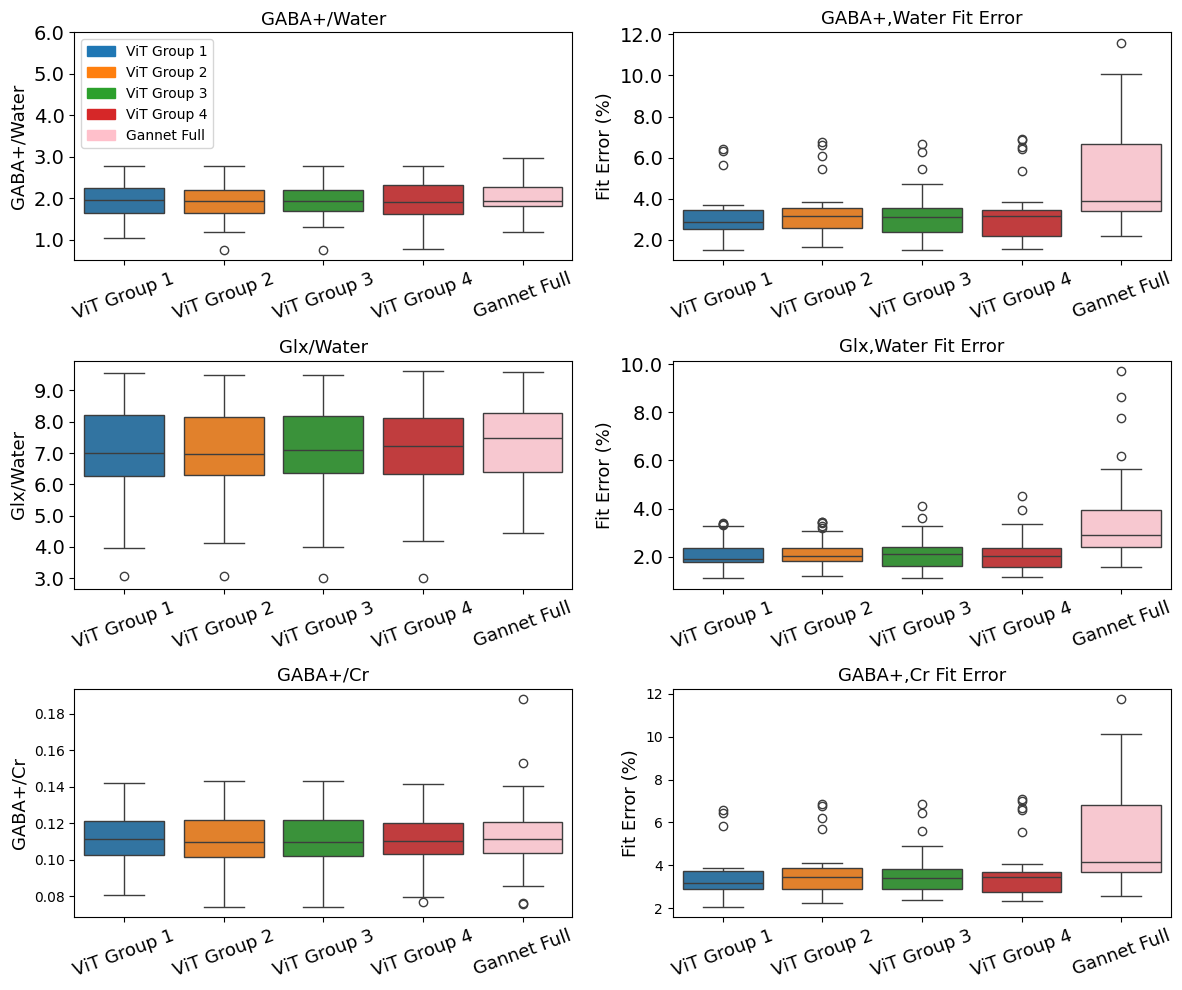}
  \caption{Box plot representations of the quantification outcomes derived from the \textit{Spectro-VIT} pipeline (proposed model) for various sets of transients are presented for the test set. Additionally, the results obtained from the \textit{Gannet-Full} pipeline, which was also applied to the test set with 320 transients, are included for reference. Transients 1-80 (ViT Group 1) are represented by the color blue, while transients 81-160 (Group 2) are depicted in orange, followed by transients 161-240 (Group 3) in green, and lastly, transients 241-320 (Group 4) in red, and the \textit{Gannet-Full} reference pipeline in pink. Overall, the outcomes of the \textit{Spectro-ViT} model showcase a remarkable resemblance to the distribution seen in the reference \textit{Gannet Full} across all groups. It is worth noting that the fitting errors in these subgroups exhibit a superior performance compared to the \textit{Gannet Full}.}\label{fig:quantification_slice_results}
\end{figure}

The investigation of the edited spectra resulting from the \textit{Spectro-ViT} pipeline for different subsets of transients demonstrates the consistency of the proposed model  (Fig.\ref{fig:metrics_boxplot_sliced} and Fig.\ref{fig:quantification_slice_results}). The results for the \textit{Spectro-ViT} Group 1 (transients 1-80) were the ones that have already been discussed previously. 

Regarding the reconstruction metrics, all of the other groups of transients achieved good results as well (Fig.\ref{fig:metrics_boxplot_sliced}). The shape score values and SNR results were found to be high, with the SNR results even surpassing those of the \textit{Gannet Full} pipeline. After an investigation of the frequency shifts across the transients from the dataset, it was shown that there is a higher susceptibility to frequency shifts for the last transients. The values of Group 4 (transients 241-320), which displayed the worst values for MSE and more dispersed values for Shape Score and SNR must be related to this result.

With respect to the quantification of the subsets, the outcomes of the \textit{Spectro-ViT} model showcase a remarkable resemblance to the distribution seen in the reference \textit{Gannet Full} across all groups, and the fitting errors in these subgroups also outperform \textit{Gannet Full}. (Fig.\ref{fig:quantification_slice_results}). The Wilcoxon signed-rank test comparing the concentration distributions of the subgroups to the reference \textit{Gannet Full} pipeline found no statistically significant differences.

\section{Discussion}

%%%%%%%%%%%% Experiment 1%%%%%%%%%%%%%%%%
%\textit{Spectro-ViT} demonstrates superior performance in reconstruction metrics. The metric in which 
\textit{Spectro-ViT } outperforms the other compared methods in all reconstruction metrics, except for SNR (Tab.\ref{table:metrics}). Notably, the fit error values of \textit{Spectro-ViT} are significantly better than those of the target, \textit{Gannet Full}, while maintaining good proximity to the curve (MSE) and similar peak shapes (shape score) to the target (Fig.\ref{fig:mixed_reconstruction} and Fig.\ref{fig:metrics_boxplot_sliced}). This is an important achievement because a lower fit error ensures more confident quantification \cite{JamieMRS}.

%The SNR values for \textit{Spectro-ViT} are also markedly superior to those of the target, with the mean value of this metric for \textit{Spectro-ViT} being almost 2.30 times higher than the mean of the same metric for \textit{Gannet Full}.

The SNR values of \textit{Spectro-ViT} surpass those of the target, with \textit{Spectro-ViT}'s mean metric value being nearly 2.30 times higher than the mean value for Gannet Full. Additionally, the SNR values for \textit{Spectro-ViT} are more consistent than those for \textit{Rdc-UNET2D} (Tab.\ref{table:metrics}).

%%%%%%%%%%%% Experiment 2%%%%%%%%%%%%%
% In terms of quantification, although Gannet-Quarter has the best values for error metrics (absolute error distribution and MAE), it exhibits a significantly larger linewidth, nearly 10\% greater than the reference, and a high standard deviation compared to other pipelines. Moreover, Gannet-Quarter's GABA+/Water fit error is almost 60\% higher than the target, with more dispersed values. Additionally, its SNR mean is almost twice as poor as the reference. Therefore, despite Gannet-Quarter's superior performance in absolute error distribution and MAE, these metrics alone do not guarantee the best quantification, as effective quantification also requires narrow linewidths \cite{choi2021spectral}, low fit error \cite{JamieMRS}, and good SNR \cite{peek2023comprehensive}. 

% This uncertainty was corroborated by the Wilcoxon signed-rank test and scatter plots of the concentrations. The Wilcoxon test indicated a statistical difference (p-value=0.0442) between the GABA+/Water concentration distributions of Gannet-Quarter and Gannet-Full. Furthermore, scatter plots suggest an overestimation bias in Gannet-Quarter's GABA concentration distributions.

In terms of quantification, although \textit{Gannet-Quarter} leads in error metrics (absolute error distribution and MAE) for quantification (Tab.\ref{table:quantification}), its spectral quality is suboptimal, as expected. It displays a linewidth nearly 10\% greater than the reference, a high standard deviation, and a GABA+/Water fit error almost 60\% higher than the target, with its SNR mean being almost twice as poor (Tab.\ref{table:metrics}). These limitations are highlighted by a Wilcoxon signed-rank test, which shows a statistical difference (p-value=0.0442) between \textit{Gannet-Quarter} and \textit{Gannet-Full} in GABA+/Water concentration distributions, and scatter plots that indicate an overestimation bias in GABA concentrations (Fig.\ref{fig:scatter_3v3_plot}). This reveals that the combination of the metrics in analysis is necessary for evaluating reconstruction \cite{berto2023advancing}. \textit{Gannet-Quarter}'s best performance in terms of absolute error distribution and MAE does not lead to the most reliable quantification. Reliable quantification requires narrow linewidths, low fit error, and good SNR \cite{choi2021spectral,JamieMRS,peek2023comprehensive}.

\textit{Spectro-ViT} boasts the best spectral quality metrics (Tab.\ref{table:metrics}) and is the only pipeline without statistically significant differences between 'predicted' concentrations and the target across all metabolites (Tab.\ref{table:quantification}). It also has the lowest coefficient of variation for both GABA+/Water and GABA+/Cr. Furthermore, scatter plot analysis reveals that, unlike \textit{Rdc-UNET2D} (which tends towards underestimation) and \textit{Gannet-Quarter} (which tends towards overestimation), \textit{Spectro-ViT} does not demonstrate such biases (Fig.\ref{fig:scatter_3v3_plot}).

These findings on \textit{Spectro-ViT}'s performance are particularly relevant in light of the study by Dziadosz et al. \cite{dziadosz2023denoising}, which revealed intrinsic significant biases and uncertainties in denoising single MRS spectra using DL. This underscores the importance of being aware of biases when developing DL-based denoising techniques for MRS. However, it is important to note that drawing general conclusions about DL denoising solely from this study is not possible. The study has limitations: it solely examined the effects of modeling on a single spectrum and relied on a limited synthetic dataset. Additionally, it utilized only two implementations of supervised DL denoising, and \textit{in-vivo} data was not incorporated.

Furthermore, the Subset Robustness experiment for \textit{Spectro-ViT} has demonstrated consistency, considering that the transients from the dataset are susceptible to different intensities of noise. The quality of all the reconstructions in this study was good distributed, comparable to the \textit{Gannet Full} (Fig. \ref{fig:metrics_boxplot_sliced} and Fig.\ref{fig:quantification_slice_results}). Furthermore, the Wilcoxon signed-rank test, which compared the concentration distributions of the subgroups to the reference \textit{Gannet Full} pipeline, found no statistically significant difference.

However, it is important to note that further studies would be necessary to enable the clinical application of this method and explore MRS distributions beyond what was used in this work in order to observe how the model will behave. Although only \textit{in-vivo} data was used in this research and the training of the model involved on-the-fly augmentations during training, it is not possible to ensure that it encompassed the MEGA-PRESS distribution for all potential sources of variance. It is necessary to test if it will also perform well with data acquired using different protocols, in smaller and more challenging brain regions, and in a broader population, including subjects suffering from diseases.

\section{Conclusions}

Experiment results demonstrate that the \textit{Spectro-ViT} model can reconstruct GABA-edited MRS spectra using only 80 transients, delivering comparable outcomes to those achieved by the standard pipeline employing the Gannet software with 320 transients. Consequently, this study's results suggest that the scans conducted herein could be expedited up to four times using this approach.

The \textit{Spectro-ViT}'s outcomes, assessed across the quality metrics MSE, SNR, Linewidth, shape score and fit error, occasionally outperform the spectral quality of the established pipeline, and the similarities in quantification distribution to the reference underscore its superior performance, surpassing the initial efforts of \textit{Rdc-UNET2D}  and securing the top-ranking position in the Edited-MRS reconstruction challenge.

The \textit{Spectro-ViT} represents a pioneering effort in MRS research, being the first to incorporate Transformer models. This innovative approach underlines the importance of exploring novel model types in advancing the field. The application of the pretrained ViT, in particular, has shown remarkable efficiency in extracting features from the time-frequency domain representation (spectrogram) of the MRS signal. Furthermore, the potential for simple modifications of the \textit{Spectro-ViT} architecture opens up new avenues for research in MRS, where the encoded representation of the spectrogram MRS signal by the ViT might go beyond reconstruction, also providing good features for classification and quantification.

In conclusion, the \textit{Spectro-ViT} model has demonstrated its potential to advance MRS research by delivering ground-breaking results in GABA-edited MRS spectra reconstruction. Its performance, surpassing that of previous models, underscores the value of embracing innovative approaches, such as Transformer models and new domains to represent the MRS signal such as spectrograms, in advancing GABA MEGA-PRESS. While these findings are promising, it is important to acknowledge the need for further research to validate the model's applicability across different acquisition protocols and a more diverse population. Additionally, more research contributions are needed to draw firmer conclusions about DL denoising in the field. These next steps will be crucial in realizing the clinical potential of the \textit{Spectro-ViT} model to accelerate the Edited-MRS scans.

% ======================================================================

%TC:ignore
% ======================================================================
\section{Acknowledgments}

The researchers from University of Campinas were supported by the DeepMind Scholarship Program, the National Council for Scientific and Technological Development (CNPq Process \#313598/2020-7), by the BI0S - Brazilian Institute of Data Science, grant \#2020/09838-0, São Paulo Research Foundation (FAPESP) - BRAINN - Brazilian Institute of Neuroscience and Neurotechnology, grant \#2013/07559-3, São Paulo Research Foundation (FAPESP), and by the Coordenação de Aperfeiçoamento de Pessoal de Nível Superior - Brasil.  The researchers from the University of Calgary were supported by NSERC Discovery Grant (\#RGPIN-2021-02867),
NSERC Discovery Grant (\#RGPIN-2017-03875), NSERC Brain CREATE Award, and Alberta Graduate Excellence Scholarship. 
% ======================================================================

% ======================================================================
% \section{References}
% ======================================================================

\bibliography{references}

% ======================================================================
%\section{Figure Captions}

\end{document}